\documentstyle[preprint,aps]{revtex}
\begin{document}
\draft
\title{\Huge Variable rest masses in 5-dimensional gravitation confronted 
with experimental data}
\author{ Luis Anchordoqui$^{\star}$, 
Graciela Birman$^{\circ\dagger}$, Santiago 
Perez 
Bergliaffa$^{\star}$\thanks{CONICET}
and H\'ector Vucetich$^{\star\dagger}$}
\address{$^{\star}$ Laboratorio de F\'\i sica Te\'orica, 
Departamento de F\'\i sica, U.N.L.P., 
c.c. 67, CP 1900,\\ La Plata, Argentina\\
$^{\circ}$ Departamento de Matem\'atica, 
Facultad de Cs. Exactas, U.N.C.P.B.A.,
Pinto 399, CP 7000, \\ Tandil, Argentina}
\maketitle
\thispagestyle{empty}
\begin{abstract}
Cosmological solutions of Einstein equation for a \mbox{5-dimensional}
space-time, in the case of a dust-filled universe, are presented. 
With these 
solutions
we are able to test a hypothetical relation between the rest mass of a 
particle and the $5^{\rm th}$ dimension. Comparison with experiment 
strongly
refutes the implied dependence of the rest mass on the cosmological 
time.
\end{abstract}
\vspace{1cm}
\newpage
\setcounter{page}{1}
The hypothesis of a higher-dimensional space-time has helped to 
formulate
theories that tackle some important problems in contemporary physics.
Kaluza--Klein theory\cite{KK}, supergravity\cite{Sg}, 
superstrings\cite{Ss},  
and extended inflation\cite{In} are but a few examples of such 
theories.
In particular, since a unified 5-dimensional theory for
electromagnetism and gravitation was developed by Kaluza\cite{Ka} and
Klein\cite{Kl}, a lot of attention has been paid to the study of 
Einstein's
equation (E. E.) in a 5-dimensional space-time. We discuss here a
completely different feature of 5-dimensional E. E. In recent years, 
some 
authors have suggested that the  
$5^ {\rm th}$ dimension might be related to mass\cite{We}. 
This idea arises either from dimensional
analysis\cite{WM,Ma}, or from reinterpretations of the 5-dimensional 
vacuum
equations\cite{Wes,Wess}. A consequence of this relation is that the 
rest 
mass of a given
body varies from point to point in space-time, in agreement with the 
ideas of
Mach\cite{Mach,Soleng}. This is a definite and testable prediction,
specially when time intervals of cosmological order are
considered\cite{We,Wesson}. With the aim of comparing it with 
experiment, we
have worked out
cosmological solutions of the 5-dimensional E. E. in the case of a 
dust-filled
universe \footnote{After this work was finished, we found a paper by
Ma\cite{Ma}, in which these cosmological solutions are worked out 
imposing 
different boundary conditions on the function $\zeta(\tau)$ 
(see below). Cosmological solutions in the case of a radiation
dominated universe were obtained by Mann and Vincent \cite{Mann}.}.

The relevant equations are \footnote{In what follows, latin indices 
$A, B,\dots$, run from 0 to 4, greek indices from 0 to 3 and latin 
indices $i,
j, \dots$, from 1 to 3.}
\begin{mathletters}
\begin{equation}
G_{AB} = 8 \; \pi \; G \; \, T_{AB}
\label{E}
\end{equation}
\begin{equation}
T_{AB} = \rho \; u_A \; u_B
\label{T}
\end{equation}
\end{mathletters}
where $\rho$ and $u_A$ are the density and the velocity field of the 
dust 
respectively, and G is Newton's gravitational constant.
\newpage

We propose the following solution 
\begin{equation}
g_{00} = \eta _{00} \; \; \; \; \; \; \; \; \; \; \; \; 
g_{ij} = \frac{A^2 (\tau)}{(1+kr^{2}/4)^2} \; \eta_{ij} \; \; \; \; \
; \; \; \; \; \; \; \; 
g_{44} = -e^{\zeta (\tau)}
\label{sol}
\end{equation}
where $r^{2}=x_1^{2}+x_2^{2}+x_3^{2}$, $\eta_{\mu\nu}$ is the 
4-dimensional 
flat metric tensor, and $k$ is the scalar curvature of the 
3-dimensional 
section (coordinates \mbox{$x_0 = x_4 = {\rm constant}$)}. The 
functions 
$A^2 (\tau)$ and $\zeta(\tau)$ need to be  determined from (1a), 
which constitutes a
system of differential equations admitting exact solutions for the 
aforementioned functions. In order to solve this system, we required
that both $\zeta$ and $ d \zeta / d \tau $ vanish at $\tau$ = today.
With these conditions, the masses of the fundamental particles can be 
set to 
their present
experimental value today\cite{pp}, and mass variations are negligible 
in short 
time scales (see eq.(\ref{masa})), which is consistent with the bound
$\vert\dot{m}/m\vert _{today}\alt 10^{-12}$ yr$^{-1}$ \cite{Pipi}. 
However, our results are completely independent of the epoch in which
the initial conditions for $\zeta$ and $d\zeta /d\tau$ were imposed. 
\footnote{This can be seen by comparing the prediction of this theory 
for
$\dot{m}/m=\dot{\zeta}/2$ at $\tau=$ today in the case where the 
initial
conditions for $\zeta$ and its derivative were imposed for instance at
the Oklo phenomenon epoch, with the current experimental limits
on $\dot{m}/m$ \cite{Pipi}.}

The solutions are explicited in the following table.

If we use them along with the definition of mass\cite{Ma} in the 
case of an
$x_4$-independent metric, 
\begin{equation}
m(\tau) = \frac{c^2}{G}\sqrt{- g^{44}} \,\,\Delta l_0
\label{masa}
\end{equation}
($\Delta l_0$ is the -finite- ``length'' of the body in the $x_4$
direction\cite{Ma}), we can now calculate the ratio 
\begin{equation}
\frac{\Delta m}{m_0}=\frac{ m (\tau_{e}) - m (\tau_{0}) } 
{ m(\tau_{0}) }
\label{coc}
\end{equation}
where $\tau_e$ is the epoch of a given event and $\tau_0$ is the
present age of the universe (see Table I).

An upper bound for $\Delta m / m_0$ can be determined with high 
accuracy at 
least for three cases: variations in the planetary paleoradius of 
Mercury
and the Moon\cite{pr}, variation of the lifetime of long-lived $\beta$
decayers (such as $^{187}$Re, $^{40}$K, and $^{87}$Rb)\cite{bd}, and 
variation in the thermal neutron capture cross section of several 
nuclear 
species in the Oklo \mbox{phenomenon\cite{ok}.} The bound for the 
mass 
variation since each event can be 
determined from the data $\dot{m}/m$ tabulated in \cite{Pipi} 
\mbox{(see
Table \ref{tabla2})}.

On the other hand, analytic expressions for the ratio $\Delta m /m_0$ 
can be calculated from Table \ref{tabla1}, (\ref{sol}), and 
(\ref{coc}) 
as functions of $\Omega_0$, $\tau_e$ and $H_0$. These expressions 
define, 
for a given $\tau_e$, 2-dimensional surfaces that intersect the plane 
$\Delta m/m_{0}=\Delta m/m_{0}\vert _{exp}$ (where 
$\Delta m/m_{0}\vert _{exp}$ is the upper bound for 
each of the events mentioned above, and is given in Table 
\ref{tabla2}), thus 
yielding 
one curve of the form $\Omega_{0}=\Omega_{0}(H_{0})$ for each event as 
follows: 
\begin{mathletters}
\begin{equation}
\Omega _{0}= \frac{1-\alpha}{(H_{0}L_{e})^2} (1- H_{0}L_{e})^2
\label{ome}
\end{equation}
\begin{equation}
\alpha = \left ( \left . 1 - \frac{\Delta m}{m_{0}} \right 
\vert _{exp}\right )
\end{equation}
\end{mathletters}              
Since this relation relation represents an upper bound for 
$\Omega _{0}$, it
implies that the allowed region for the theory in the $(H_0, 
\Omega _0)$
plane lies below the curve (\ref{ome}).

Besides, the density $\rho(\tau_{0},
H_{0})$ (and then $\Omega_{0}(\tau_{0},H_{0}$)) can also be calculated
from (\ref{E}) and (\ref{sol}). If we take $\tau_0$=10 Gyr (20 Gyr) 
as a 
lower (upper) bound for the present age of the Universe \cite{drunk}, 
the latter 
$\Omega_0$ curves define a strip in the $H_{0}-\Omega_{0}$ plane. The 
intersection of this strip with the window determined by the present 
experimental limits for $\Omega_0$ and $H_0$ \footnote{We took 
$0.15\alt
\Omega_{0}\alt 2$ \cite{t1}, and $0.4\alt h\alt 1$ \cite{kt}, with 
$H_{0}=100\;h$ km seg$^{-1}$ Mpc$^{-1}$ as usual.} and with the region 
$\Omega_{0}>1$ ($\Omega_{0}<1$) for the spherical (pseudospherical) 
case
is the region 
of validity of the theory in the $H_{0}-\Omega_{0}$ plane. That is to 
say,
all the predictions of the theory for $\Omega_0$ should fall in this 
region
(which we call ${\cal R}$) for the theory to be valid.

Our results are shown in Fig. 1. It can be easily seen that there 
is a strong disagreement between theory and observation: the curves 
for the long-lived $\beta$ decayers and the lunar paleoradius do not
intersect the region ${\cal R}$, and this is true no matter which 
3-geometry 
(plane, spherical, or
pseudospherical) one considers. The Oklo phenomenon results 
exclude the 
closed model and flat models but are partially consistent 
with an open  Universe.

We are thus led to conclude that the 
assumed relation connecting the $5^{\rm th}$ dimension and mass is 
false
at least for the case of the generalized metric given by (\ref{sol}). 
More general ({\rm e.g.} $x_4$-dependent) metrics will be 
studied elsewhere.

\begin{table}
\caption{Cosmological solutions of E. E.}
\begin{tabular}{cccc}

 & $k \neq 0$ & $k = 0$  \\ 
\hline
 $A^2 (\tau) / c^2 $ & $\tau (-k\tau + 2 \tilde{\tau})$ 
&$2 \tau \tilde{\tau}$ 
\\  \hline
$\tilde{\tau}$ &  $k\tau_0 \frac{\tau_0 H_0 - 1}{ 2 \tau_0
H_0 - 1}$ & -  \\ \hline
$\zeta (\tau)$& $ \ln \left[ \frac{ ( H_{0} \tau + 1 - H_{0} 
\tau_{0} )
^{2} 
( 2 \tilde{\tau}
\tau_{0} - k \tau_{0}^{2})}{ 2 \tilde{\tau} \tau - k \tau^{2}}
\right] $  
&$ \ln \left[ \frac{1}{2} \sqrt{\frac{\tau}{\tau_{0}}} + 
\frac{1}{2} \sqrt{\frac{\tau_{0}}{\tau}}
\;\right]^{2}  $  \\ \hline
$\rho(\tau_{0},H_{0})$ & $ \frac{3}{8 \pi G} 
\frac{( 1 - \tau_{0} H_{0} )^{2}}{
\tau_{0}^{2}}$ & $ \frac{3 H_{0}^{2}}{8 \pi G}$ \\ \hline
$\Omega (\tau_0,H_{0})$ &  $\left( \frac{1}{\tau_{0} H_{0}} - 
1 \right) ^{2}$
& 1   \\ \hline
$\frac{\Delta m}{m}(H_{0},\Omega _{0})$ & $ 1 - 
\sqrt{\frac{[ 1 - L_{e} H_{0} (1 + \surd 
\bar{\Omega_{0}}) ] \, 
[ 1 + L_{e} H_{0}
( \surd \bar{\Omega_{0}} - 1 ) ]}{( 1 - H_{0} L_{e} )^{2}}}$ & 
 $1 - \sqrt{\frac{ 1 - 2 L_{e} H_{0}} {( 1 - H_{0} L_{e} )^{2}}}$ 
\end{tabular}
\label{tabla1}

\vspace{.5cm}

( $H_0$ is the value of the Hubble constant today, $\Omega _0$ 
is the ratio
between the present and critical density for the universe and 
$L_e$ is the
age of the event $\tau_0 - \tau_e$.)
\end{table}

\begin{table}
\caption{Sample experimental data from $\left .
\frac{\Delta m}{m}\right
\vert _{exp}$}
\begin{tabular}{cccc}

   & Value [ $10^{-11}$ yr$^{-1}$ ] &
Age [Gyr] & $\left .\frac{\Delta m}{m}\right\vert _{exp}$\\ \hline
Planetary paleoradius& & & \\ \hline 
Moon &  $\frac{\Delta R}{R\Delta t}\leq$ 0.015 & 3.9 & 6.5$ 
\times 10^{-4}$
\\ \hline
Long-live $\beta$ decayers& & & \\ \hline 
$^{40}$K & $\frac{\Delta\lambda}{\lambda\Delta t}\leq$ 
0.29 & 4.6 & 2.3 
$\times 10^{-4}$\\ \hline
Oklo phenomenon& & & \\ \hline
$^{157}$Gd & $\frac{\Delta\sigma}{\sigma\Delta t}\leq$ 123 
& 1.8 & 3.3 
$\times 10^{-3}$ \\ 
\end{tabular}
\label{tabla2}
\end{table}


\begin{references}
\bibitem{KK} For an update, see ``Kaluza-Klein theory in 
perspective'', M.J.
Duff, talk delivered at the Oskar Klein Centenary Nobel 
Symposium, Stockholm
(1994). See also: A. Chodos and S. Detweiler, Phys. Rev. D {\bf21},
2167 (1980); P. G. O. Freund, Nucl. Phys. B {\bf 209}, 146 (1982).
\bibitem{Sg} P. Van Niewenhuizen, Phys. Reports {\bf
68}, 189 (1981).
\bibitem{Ss} M. B. Green, J. H. Schwartz and E. Witten, ``Superstring
theory''. Cambridge monographs on mathematical physics. Cambridge
University Press (1987).
\bibitem{In} A.S. Majumdar and S.K. Sethi, Phys. Rev. D {\bf 46}, 
5315 (1992).
\bibitem{Ka} T. Kaluza, Sitz. Preus. Akad. Wiss. {\bf 33}, 966 (1921).
\bibitem{Kl} O. Klein, Z. Phys. {\bf 37}, 895 (1926).
\bibitem{We} P. Wesson, Gen. Rel. Grav {\bf 16}, 193 (1984).
\bibitem{WM} B. Mashhoon, H. Liu, P. Wesson, Phys. Lett. B. 
{\bf 331}, 305 (1994).
\bibitem{Ma} G. Ma, Phys. Lett. A. {\bf 143}, 183 (1990).
\bibitem{Wes} P. Wesson, Mod. Phys. Lett. A. {\bf 7}, 921 (1992).
\bibitem{Wess} P. Wesson, J. Ponce de Leon, J. Math. Phys. 
{\bf 33}, 3883 (1992).
\bibitem{Mach} E.Mach,``The Science of Mechanics'', 2$^{\rm nd}$ 
edition, 
Open Court
Publishing Co., La Salle, Illinois (1983).
\bibitem{Soleng} It should be stressed that such ideas raise some
important conceptual problems. See the discussion in \O. Gr{\o}n 
and H. Soleng, Gen. Rel. Grav {\bf 20}, 1115 (1988).
\bibitem{Wesson} P. S. Wesson, em Gen. Rel. Grav. {\bf 22}, 
707 (1990).
\bibitem{Mann} R. B. Mann and D. E. Vincent,  Phys. Lett. A {\bf
107}, 75 (1985).
\bibitem{pp} Review of Particle Propieties {\em Phys. Rev.} D. 
{\bf 45},
S1 (1992).
\bibitem{Pipi} P. Sisterna and H. Vucetich, Phys. Rev. D. {\bf 44}, 
3096 (1991).
\bibitem{pr} M.W, McElhimmy, S.R. Taylor, amd D. Stephenson, 
Nature (London)
{\bf 271}, 316 (1978).
\bibitem{bd} G.W. Weterhill, Ann. Rev. Nucl. Sci., {\bf 25}, 283 
(1975).
\bibitem {ok} A.I. Shlyakhter, Nature (London) {\bf 264}, 340 (1976).
\bibitem{drunk} P. Coles, G. Ellis, {\em Nature} {\bf 370}, 609 
(1994).
\bibitem{t1} M. S. Turner, talk presented at the NAS Special 
Colloquium on 
Physical Cosmology, Irvine, March 1992.
\bibitem{kt} E. W. Kolb, M. S. Turner, ``The Early Universe''
(Addison-Wesley Publishing Company, 1990).
\end{references}
\end{document}